\documentclass{article}

\usepackage[preprint,nonatbib]{nips_2018}

\usepackage[utf8x]{inputenc} 
\usepackage[T1]{fontenc}    
\usepackage{hyperref}       
\usepackage{url}            
\usepackage{booktabs}       
\usepackage{amsfonts}       
\usepackage{nicefrac}       
\usepackage{microtype}      
\usepackage{float}
\usepackage[hang,flushmargin]{footmisc}
\usepackage{graphicx}
\usepackage{subfigure}
\usepackage{amssymb}
\usepackage{mathtools}
\usepackage{booktabs}
\usepackage{enumitem}
\usepackage{makecell}
\usepackage{pbox}
\usepackage{amsmath}
\usepackage[vlined,ruled,linesnumbered]{algorithm2e}

\usepackage{color}

\title{A Fully Private Pipeline for Deep Learning on Electronic Health Records}

\author{
Edward Chou$^1$\thanks{authors contributed equally}\qquad
Thao Nguyen$^1$\footnotemark[1] \qquad
Josh Beal$^1$ \qquad
\textbf{Albert Haque$^1$ \qquad
Li Fei-Fei$^1$} \\
\\
$^1$Department of Computer Science, Stanford University
}

\begin{document}

\maketitle

\begin{abstract}

We introduce an end-to-end private deep learning framework, applied to the task of predicting 30-day readmission from electronic health records. By using differential privacy during training and homomorphic encryption during inference, we demonstrate that our proposed pipeline could maintain high performance while providing robust privacy guarantees against information leak from data transmission or attacks against the model. We also explore several techniques to address the privacy-utility trade-off in deploying neural networks with privacy mechanisms, improving the accuracy of differentially-private training and the computation cost of encrypted operations using ideas from both machine learning and cryptography.

\end{abstract}


\section{Introduction}\label{sec:intro}

Deep neural networks have been applied to a variety of clinical tasks to great success. Medical imaging diagnosis \cite{gulshan2016development}, genome processing \cite{nguyen2017metagenomic}, and disease onset predictions \cite{liu2018deepehr} are domains where deep learning could help uncover patterns in data and greatly improve quality of care and treatment. Unfortunately, since medical data is also extremely privacy-sensitive, the healthcare industry is subject to stringent patient protection regulations such as HIPAA and GINA, impeding the widespread adoption of data mining techniques in the medical community \cite{asghar2017hipaaprivacyhealth}. 


Without addressing these privacy concerns, it is unlikely that machine learning as a service (MLaaS) platforms will be adopted by the healthcare community due to the risk of information leakage during data transmission or to cloud providers \cite{bae2018securityprivacyissues}. Anonymized data is vulnerable to de-anonymization attacks, as shown in the Netflix deanonymization demonstration \cite{narayanan2008netflixdeanonymization}, and thus is not an adequate solution for data sharing in the healthcare domain \cite{gymrek2013identifyinggenomesurname}. The alternative scheme of deploying only the trained models is also insufficient, as recent works \cite{yeom2017overfittingMI, carlini2018secretsharer} have demonstrated how neural networks could memorize training data even when they are not overfitting.  Attacks like membership inference \cite{shokri2016membershipinference} and model inversion \cite{fredrickson2015modelinversion} can reveal population information or recover training inputs from a neural network, in some cases with only black-box access to the model.

\begin{figure}[thbp]
\centering
\includegraphics[width=4.85in]{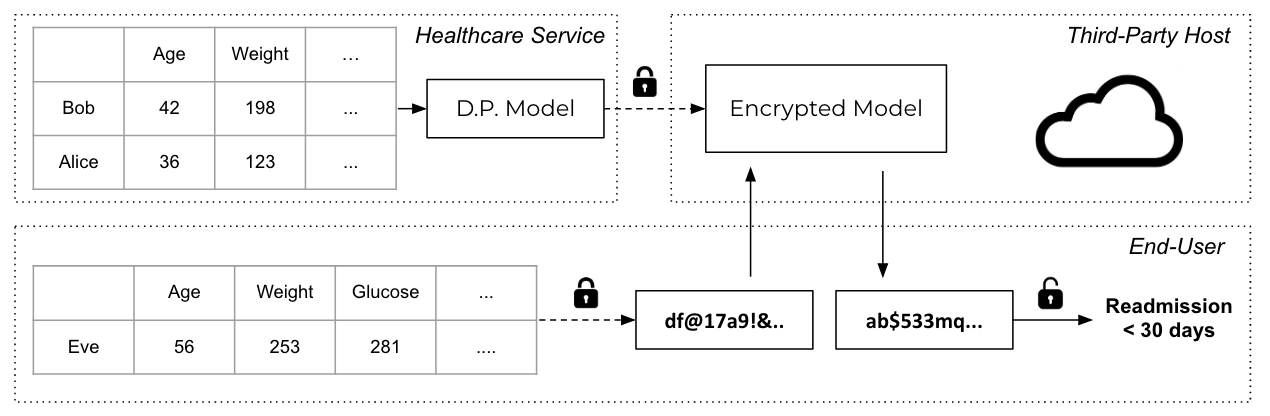}
\caption{The healthcare provider trains a neural network using differential privacy and hosts the encrypted model on a third-party host, allowing end users to send and receive encrypted data. }
\label{fig:method}
\end{figure}

Machine-learning services that are private and secure by design will allow healthcare practitioners to benefit from advances in deep learning.  In this work, we explore separate constructs for private and secure machine learning that are compatible, using differential privacy during training \cite{abadi2017privatetwoapproaches} and homomorphic encryption for inference \cite{gilad2016cryptonets}, in order to provide a fully-private pipeline. 
In addition to our work being the first to combine and apply these techniques to realistic clinical tasks, we also propose several guidelines to improve the accuracy and computational performance. In particular, we demonstrate the importance of standardizing EHR data to enhance differentially private learning in terms of privacy costs and training stability.
We also use state-of-the-art techniques to improve network performance and computational overhead through parameterized activation functions with coefficients quantized to leverage sparse polynomial multiplication.

\section{Related Work}
Differential privacy is a privacy construct which guarantees that an individual will not change the overall statistics of the population \cite{dwork2008differential}, a formal definition defined as algorithm M and dataset D being $(\epsilon,\delta)$ private if $P(M(x \in D) \in S)\leq e^\epsilon P(M(x \in D')\in S) + \delta$. Applying differential privacy to neural networks helps ensure defenses against membership inference and model inversion attacks \cite{abadi2017privatetwoapproaches}. This can be achieved by either applying noise to gradients while training a single model \cite{abadi2016dpsgd, song2013stochastic} or by segregating data and adding noise in a collaborative learning setting \cite{papernot2018scalable, shokri2015privacy}. 

Gentry et. al. \cite{gentry2009fully} introduced fully homomorphic encryption (FHE) which allows anyone to perform computation over encrypted data without having to decrypt it.
A weaker version of FHE, called leveled homomorphic encryption (LHE) permits a subset of arithmetic operations on a depth-bounded arithmetic circuit \cite{brakerski2014efficient}.
CryptoNets \cite{gilad2016cryptonets} was one of the first works to apply LHE to a neural network setting.
More recently, \cite{chabanne2017privacy} and \cite{hesamifard2017cryptodl} extended this technique to deeper network architectures and developed low-degree polynomial approximations to common activation functions (i.e. ReLU, Sigmoid), in addition to leveraging batch normalization for stability.  

Several previous works have attempted to apply privacy techniques to the healthcare setting. \cite{joppe2014encryptedmedical} uses homomorphic encryption to encrypt a linear regression model trained on medical databases.  A good deal of literature also studies the use of differential privacy (DP) in medicine \cite{shaked2016publishingdpmedical, dankar2012dphealth, dankar2013dphealthcare}, although the focus is mainly on applying DP to the datasets rather than the ML algorithm.

\section{Methods}

\subsection{Differentially Private Stochastic Gradient Descent (DP-SGD)}
DP-SGD optimization was developed by \cite{abadi2016dpsgd} and involves adding Gaussian noise and clipping gradients of neural networks during training with stochastic gradient descent. It also keeps track of the privacy loss through a privacy accountant \cite{mcsherry2009privacyaccounting}, which prematurely terminates training when the total privacy cost of accessing training data exceeds a predetermined budget. Differential privacy is attained as clipping bounds the L2-norm of individual gradients, thus limiting the influence of each example on the learning updates.  An outline of DP-SGD algorithm is included in the appendix.


Through standardization, we scale and translate each feature such that its values lie between 0 and 1. As seen from Table \ref{tab:standardization_results}, this greatly reduces the L2 norm of the gradients, and equivalently, the clipping bound and the amount of noise required for privacy guarantees. We also observe that standardization improves both AUC and recall, which are especially important given the scarcity of positive labels.

\subsection{CryptoNets - Inference on Encrypted Data}
We use a levelled HE-scheme with a pretrained network as outlined in \cite{gilad2016cryptonets} to support inference for encrypted input. We use the FV-RNS scheme proposed by \cite{bajard2016full}. This is a residue number system (RNS) variant of the FV encryption scheme \cite{fan2012somewhat} and is implemented in SEAL, a library for homomorphic encryption \cite{seal23}. We use a ring dimension $n = 8192$ with two plaintext moduli $t^{(j)}$. Each coefficient modulus $q^{(j)}$ is decomposed into four 64-bit moduli for efficient use of FV-RNS. Further details of the encryption operation are placed in the appendix in the interest of space.


Neural networks mostly consist of HE compatible multiplicative and additive operations, with the exception of non-linear activation functions which the original CryptoNets paper \cite{gilad2016cryptonets} substitutes with a square activation. However, the activation function of a neural network is critical for convergence \cite{glorot2011deep}, and it has been shown that polynomial approximations of activation functions retain much of the performance as their nonlinear counterparts \cite{gautier2016globally, livni2014computational}.  We polynomially approximate the Swish activation using a Minimax function (details provided in appendix), giving us a polynomial equation $p = 0.12050344x^2 +  0.5x + 0.153613744$.  Due to the extra multiplicative operations which are expensive in HE schemes, using more complex polynomial functions requires more computational power.  However, while a brute-force implementation would require $\mathcal{O}(n^2)$ time to complete, HE methods are able to accomplish this in $\mathcal{O}(n\log{}n)$ when a coefficient modulus $q$ is chosen such that $(q-1)$ is divisible by $2n$, by the Number Theoretic Transform \cite{harvey2014faster}.  Thus, we use the activation $p^* = 2^{-3}x^2 + 2^{-1}x + 2^{-4}$ which helps lower the computational cost of our chosen activation layer.

\section{Experiments}
\subsection{Dataset and Model}
Our dataset \cite{strack2014impact} is obtained from the UCI Machine Learning Repository and contains 10 years (1999-2008) of medical records of more than 101,000 patients from 130 US hospitals. The data consists of demographics and clinical metrics associated with risk of diabetes, in addition to readmission outcome. Features with about 40\% missing values such as medical speciality, payer code and weight are removed from our analysis. We aggregated ICD9 codes that represent similar diagnoses into 10 groups, and converted any categorical feature into one-hot encoding representation. Then we randomly split the dataset into train and test sets with the ratio of 75:25. Our goal is to predict whether a diabetic patient would be readmitted within 30 days after being discharged. Being a key indicator of quality of care, this task has been widely studied in existing literature, and for this dataset, by other works such as \cite{chopra2017recurrent} and \cite{bhuvan2016ehrdiabetic}. Our network consists of one hidden layer of size 32 and one output layer, each is followed by an approximated quantized swish activation function. We also use a mean squared error weighted by the class imbalance ratio (8:1). The model was trained with batch size of 256 and Adam optimizer on an Intel Core i7-5930K CPU at 3.50GHz with 48GB RAM.

\subsection{Prediction performance}
\begin{table*}[tbp]
    \centering
    \begin{tabular}{|c|c|c|c|c|c|c} \hline
    DP noise injected & Standardization & Median L2-Norm & Accuracy & AUC & Recall\\ \hline
    No & No & 50.6 & \textbf{0.802} & 0.659 & 0.313 \\
    No & Yes & \textbf{2.34} & 0.610 & \textbf{0.677} & 0.642 \\
    Yes & No & N/A & 0.765 & 0.615 & 0.356 \\
    Yes & Yes & N/A & 0.588 & 0.638 & \textbf{0.662} \\
   \hline
    \end{tabular}
    \label{tab:standardization_results}
    \caption{Effects of input standardization and DP training (with small noise $\epsilon=8$) on different metrics.  L2-Norm is much smaller with standardization (results not included for DP-noise due to clipping).  We find that overall DP does not hurt the AUC much and even improves the recall of our experiments.}
\end{table*}

\begin{figure}[ht]
\centering
\includegraphics[width=2.0in]{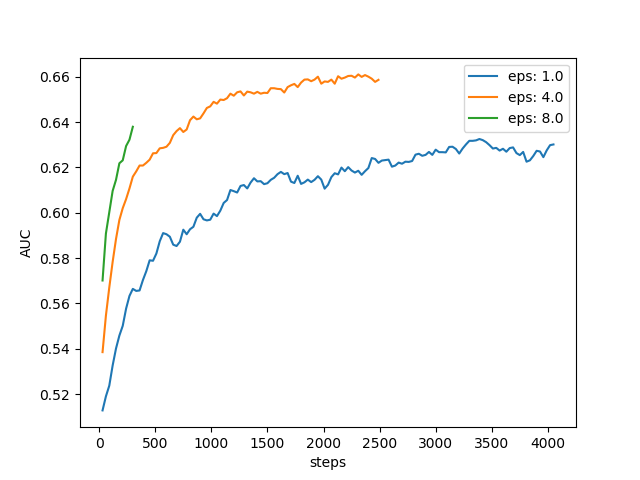}
\includegraphics[width=2.0in]{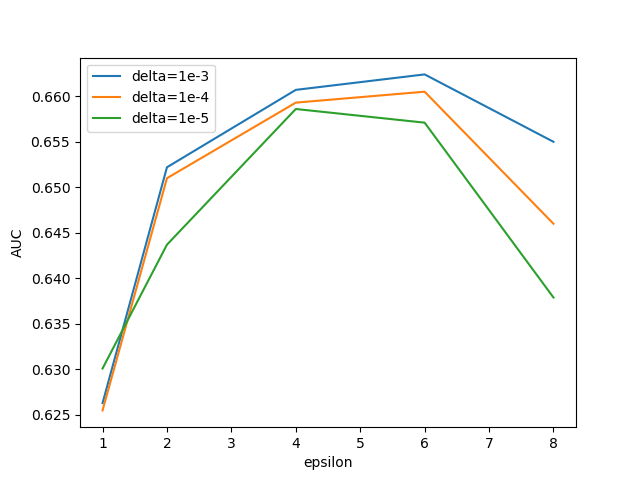}
\caption{Left: AUC over time for different noise levels.  We note that the budget $\delta=1e-5$ is spent faster for smaller noise (i.e. larger eps) leading to early termination. Right : AUC of different $(\epsilon, \delta)$ values on test set.  Both plots show that the best AUC is achieved with a moderate noise level: when $\epsilon=4$, we obtain AUC = 0.66, recall = 0.60}
\label{fig:perf}
\end{figure}

We fix to $\delta=10^{-5}$ by estimating $\delta=1/n$ where $n=100,000$ rows, and clipping bound $=1.0$ using the l2-norm estimation. We report test accuracy, area under Receiver Operating Characteristic curve (AUC) and recall of our model's predictions on test dataset, with more emphasis on the last 2 metrics due to the imbalanced nature of the training data. AUC measures the discrimination at different classification thresholds, while high recall is necessary as we consider the cost of misses (i.e. discharging patients when they are not ready) to be more serious that that of false alarms. In fact, we find high accuracy to often be correlated with poor performance, as it is trivial to get ${\sim}$90\% accuracy by guessing one class. From Table \ref{tab:standardization_results}, the standard training of our HE-compatible model (with polynomial activations) gives better performance (AUC$=0.67$) than the baselines obtained from normal neural networks in \cite{chopra2017recurrent} (AUC $=0.61$) and \cite{bhuvan2016ehrdiabetic} (AUC $=0.23$). Even with large amount of noise injected (eps=1.0 in Figure \ref{fig:perf}), our fully private network still yields a higher AUC of 0.63, compared to the aforementioned existing works on this dataset. We found that with $\delta=10^{-5}$, the best performance is achieved by injecting a moderate level of noise ($\epsilon=4$).



\subsection{Comparison of Activation Functions}
Before approximated swish activation, we also experimented with other functions, both low-degree polynomial (square activation) and non-linear (ReLU and Sigmoid). Test results from training our network with each function combination with no differential private noise are shown in Table \ref{tab:activation_results}, with confidence level obtained from 10 repeated trials. We observe that square activations produce more instability in performance and high variance across different runs. For approximated swish activations, the AUC is fairly stable and even surpasses the performance of non-linear activations. 

\begin{table*}[ht]
    \centering
    \begin{tabular}{|c|c|c|c|c} \hline
     Activations & Accuracy & AUC & Recall\\ \hline
     Square$\times{2}$ & \textbf{0.650} $\pm$ \textbf{0.141} & 0.654 $\pm$ 0.014 & 0.552 $\pm$ 0.167 \\
     ReLU-Sigmoid & 0.633 $\pm$ 0.125 & 0.668 $\pm$ 0.003 & 0.596 $\pm$ 0.163 \\
     Approximated Swish$\times{2}$ & 0.618 $\pm$ 0.153 & \textbf{0.678} $\pm$ \textbf{0.003} & \textbf{0.645} $\pm$ \textbf{0.181}
    \\  \hline
    \end{tabular}
    \label{tab:activation_results}
    \caption{Accuracy of Activation Functions.  We see that the square activation produces the lowest AU and recall.  We also find the approximated swish function produces higher AUC and recall values than the nonlinear ReLU-Sigmoid activations.}
\end{table*}


\begin{table*}[!htb]
    \centering
    \begin{tabular}{|c|c|c|c|c} \hline
     Activations & Wallclock Runtime (s) & Multiplicative Operations\\ \hline
     Square & 21.700657 & 6780 \\
     Approx. Swish (without quantization) & 21.900602 & 6912  \\
     Approx. Swish (with quantization) & 21.797248 & 6912
    \\  \hline
    \end{tabular}
    \label{tab:computational_costs}
        \caption{Computational Costs of Activation Functions: We can see that the non-quantized swish has longer runtime due to the higher amount of multiplicative operations, and also that the quantized swish approximation reduces the runtime without reducing the number of operations.}
\end{table*}

Homomorphic encryption is computationally expensive, raising the runtime of neural network inferences from an order of milliseconds to $\approx$20 seconds.  We can see in Table \ref{tab:computational_costs} that the approximated swish function without quantization adds additional runtime due to the extra multiplicative operations, but with quantization, the increased costs become negligible.  With larger networks that contain more activations and multiplicative operations, this time saving will be even more pronounced.
\section{Conclusion}

For deep learning to be widely adopted in the healthcare community, the privacy and security of patient data will have to be ensured at every step of deployment.  We utilized differentially private learning and homomorphic encryption to protect privacy at both the training and inference stages, and demonstrated the deployment of our framework on a representative clinical prediction task. We also discussed several techniques to minimize the training instability and computational trade-off incurred by those privacy measures. We hope this work will inspire future efforts to build machine learning systems that prioritizes patient privacy by design.

\clearpage
\newpage
\small
\bibliographystyle{abbrv}

\normalsize
\clearpage
\newpage

\appendix 

\clearpage
\newpage

\onecolumn

\section*{Appendix}
\appendix
\section{Differentially Private Gradient Optimization}
\begin{algorithm}
\SetKwData{Left}{left}\SetKwData{This}{this}\SetKwData{Up}{up}
\SetKwFunction{Union}{Union}\SetKwFunction{FindCompress}{FindCompress}
\SetKwInOut{Input}{input}\SetKwInOut{Output}{output}
\Input{ Examples $\{x_1, . . . , x_N \}$, loss function $\mathcal{L}(\theta) = \frac{1}{N} \Sigma_i \mathcal{L} (\theta, x_i)$, Parameters: learning rate $\eta_t$, noise scale $\sigma$, group size $L$, gradient norm bound $C$.}
\Output{$\theta_T$ and calculate privacy cost $(\epsilon, \delta)$ using a privacy accountant method}
\textbf{Initialize} $\theta_0$ randomly\;
\For{$t \in [T]$}{
Take a random sample $L_t$ with sampling probability $L/N$\;
\textbf{Compute gradient}\;
For each $i \in L_t$, compute $\textbf{g}_t(x_i) \leftarrow \nabla_{\theta_t} \mathcal{L}(\theta_t, x_i)$\;
\textbf{Clip gradient}\;
$\overline{\textbf{g}}_t(x_i) \leftarrow \textbf{g}_t (x_i) / \text{max}(1, \frac{\lVert \textbf{g}_t (x_i) \rVert_2}{C}$\;
\textbf{Add noise}\;
$\Tilde{\textbf{g}}_t \leftarrow \frac{1}{L} (\Sigma_i (\overline{\textbf{g}}_t(x_i) + \mathcal{N}(0, \sigma^2 C^2 \mathcal{I})))$\;
\textbf{Descent}\;
$\theta_{t+1} \leftarrow \theta_t - \eta_t \Tilde{\textbf{g}}_t$\;
}
\caption{Differentially private SGD}\label{algo_dpgradopt}
\end{algorithm}\DecMargin{1em}

\section{Levelled Homomorphic Encryption}
The levelled homomorphic encryption scheme is a structure-preserving transformation between two algebraic structures, which can be leveraged by cryptosystems to allow for arithmetic operations on encrypted data.
Let $R_k$ denote the polynomial ring $\mathbb{Z}_k[x]/(x^n + 1)$.
We let $x \leftarrow S$ denote uniformly random sampling of $x$ from an arbitrary set $S$, and $\lfloor \frac{t}{q} p \rceil$ denote a coefficient-wise division and rounding of the polynomial $p$ with respect to integer moduli $t$ and $q$.
Let $[p]_q$ denote the reduction of the coefficients of the polynomial $p$ modulo $q$, and let $\Delta$ denote $\lfloor {q/t} \rfloor$.

\textbf{Encryption Scheme.}
Bajard et al. \cite{bajard2016full} proposed an encryption scheme, FV-RNS, which is a residue number system (RNS) variant of the FV encryption scheme. 
In FV-RNS, plaintexts are elements of the polynomial ring $R_t$, where $t$ is the plaintext modulus and $n$ is the maximum degree of the polynomial, which is commonly selected to be one of $\{1024, 2048, 4096, 8192, 16384, 32768\}$. 
The plaintext elements are mapped to multiple ciphertexts in $R_q$ in the encryption scheme, with $q \gg t$ as the ciphertext coefficient modulus.
For any logarithm base $\beta$, let $\ell = \lfloor {\log_\beta q} \rfloor$ be the number of terms in the base-$\beta$ decomposition of polynomials in $R_q$ that is used for relinearization.

Let $\chi$ denote the truncated discrete Gaussian distribution.
The secret key is generated as $s \leftarrow R_3$ with coefficients $s_i \in \{ 0, 1, -1\}$.
The public key $(p_0, p_1)$ is generated by sampling $p_0 \leftarrow R_q$ and $e' \leftarrow \chi$ and constructing $p_1 = [−(s p_0 + e')]_q$.
The evaluation keys $(a_i, g_i)$ are generated by sampling $a_i \leftarrow R_q$ and constructing $g_i=[−(a_i s + ie') + \beta^i s^2]_q$ for each $i \in \{0, ..., \ell\}$.

A plaintext $m \in R_t$ is encrypted by sampling $u \leftarrow R_3$ with coefficients $u_i \in \{ 0, 1, -1\}$ and $e_1,e_2 \leftarrow \chi$, and letting $(c_0,c_1)=([\lfloor {q/t} \rfloor m+p_0u+e_1]_q,[p_1u+e_1]_q)$.
A ciphertext $(c_0,c_1) \in R_q \times R_q$ is decrypted as $m=[\lfloor \frac{t}{q} [c_0+c_1s]_q \rceil]_t \in R_t$.

\textbf{Arithmetic}.
The addition of two ciphertexts $(c_0,c_1)$ and $(d_0,d_1)$ is $(c_0+d_0,c_1+d_1)$.
The multiplication of two ciphertexts $(c_0,c_1)$ and $(d_0,d_1)$ occurs by constructing
\begin{equation*}
    c_0' = \left[\left\lfloor \frac{t}{q} [c_0 d_0] \right\rceil\right]_q \textrm{ ,\quad } c_1' = \left[\left\lfloor \frac{t}{q} [c_0 d_1 + c_1 d_0] \right\rceil\right]_q, \textrm{ and \,} c_2' = \left[\left\lfloor \frac{t}{q} [c_1 d_1] \right\rceil\right]_q .
\end{equation*}
We express $c_2'$ in base $\beta$ as $c_2' = \sum_{i=0}^{\ell} c_2'^{(i)} \beta^i$. We then let $r_0=c_0'+\sum_{i=0}^{\ell} a_i c_2'^{(i)}$ and $r_1= c_1'+\sum_{i=0}^{\ell} g_i c_2'^{(i)}$, which forms the product ciphertext $(r_0, r_1) \in R_q \times R_q$.

The addition of ciphertext $(c_0,c_1)$ and plaintext $m$ is the ciphertext $(c_0+\Delta m,c_1)$. The multiplication of ciphertext $(c_0,c_1)$ and plaintext $m$ is the ciphertext $(mc_0,mc_1)$.

The advantage of the residue number system variant is that the coefficient modulus $q$ can be decomposed into several small moduli $q_1, ..., q_k$ to avoid multiple-precision operations on the polynomial coefficients in the homomorphic operations, which improves the efficiency of evaluation.

\textbf{Integer Encoder}.
To encode real numbers involved in the computation, we choose a fixed precision for the values (15 bits) and scale each value by the corresponding power of 2 to get an integer for use with the encoder described below. After decryption, we can divide by the accumulated scaling factor to obtain a real value for the prediction.
The encoder consists of a base-2 integer encoder \cite{seal23}.
For a given integer $z$, consider the binary expansion of $|z| = z_{n-1}...z_1 z_0$.
The the coefficients $b_i$ of the polynomial $f(x) = \sum_{i=0}^{n - 1} b_ix^i$ in the plaintext ring are $z_i$ if $z_i \geq 0$ otherwise $b_i=t-z_i$.


\textbf{Polynomials} 
Let $x \in \mathbb{R}$ and let $f: \mathbb{R} \rightarrow \mathbb{R}$ denote the activation function.
Our task is to approximate $f$ with a polynomial $p^*: \mathbb{R} \rightarrow \mathbb{R}$ where $p^*(x) = p^*_0 + p_1^* x + \cdots + p^*_n x^n$ subject to the constraint that each coefficient is a power of 2.
Define $\mathcal{P}^{(2)}_n$ as the set of all polynomials of degree less than or equal to $n$, such that all coefficients are base-2.
That is, $\mathcal{P}^{(2)}_n = \{  2^{a_0} + 2^{a_1} x + \cdots + 2^{a_n} x^n , a_i \in \mathbb{Z} \}$.
Let $p$ be the minimax approximation to $f$ on some interval $[-a, a]$.
Let $\hat{p}$ be the same as $p$, but with all coefficients rounded to the nearest $2^k$ where $k \in \mathbb{Z}$.
Note, $\hat{p} \in \mathcal{P}^{(2)}_n$.

\textbf{Maximum Error \& Minimax} 
The maximum difference (i.e., error) $\delta$ between two functions $g$ and $h$ is $\delta(g, h) = \max_{x\in [-a,a]} | g(x) - h(x) |$.
This provides a strong bound on the optimal polynomial approximation error $\delta(f, p^*)$ where $\delta(f, p) \leq \delta(f, p^*) \leq \delta(f, \hat{p}).$
We state minimax problem as follows. For a given activation function $f$, we seek to find the best polynomial $p^* \in \mathcal{P}^{(2)}_n$ such that,
\begin{equation}
    \delta(f, p^*) = \min_{q \in \mathcal{P}^{(2)}_n} \delta(f, q)
\end{equation}
subject to the constraint,
\begin{equation}\label{eq:constraint}
    \delta(f, p^*) \leq K \textrm{ where } K \geq \delta(f, p).
\end{equation}

\textbf{Finite Number of Solutions.}
Let $d, n \in \mathbb{N}$, and $D = \{0,...,d\}$. For $\forall i \in D$, let $x_i, l_i, u_i \in \mathbb{R}$ such that $x_j \neq x_k$ if $j\neq k$. We can construct a bounded polyhedron,
\begin{equation*}
    \mathcal{B} = \left\{ (\alpha_0, ..., \alpha_n) \in \mathbb{R}^{n+1} \; \Bigg\vert \; l_i \leq \sum\limits_{j=0}^{n} \alpha_j x^j_i \leq u_i, \forall i \in D  \right\}
\end{equation*}
where each $(\alpha_0,...,\alpha_n)$ tuple represents any polynomial $q \in \mathcal{P}^{(2)}_n$, and where $\alpha_i$ represents the degree $i$ coefficient.
\cite{brisebarre2006computing} show that the number of polynomials satisfying Equation \ref{eq:constraint} is finite if the polynomials are contained in $\mathcal{B}$.
They also proposed an efficient scanning method to find the optimal polynomial approximation $p^*$.
Equipped with our new-found approximation $p^*$, we can evaluate the effectiveness of $p^*$ as an activation function in both non-encrypted and encrypted domains.

\end{document}